\documentclass[12pt]{article}

\usepackage[cp1251]{inputenc}

\input{epsf}

\usepackage[pdftex]{graphicx}

\begin{document}




\author{  G.~G.~Kozlov,  V.~S.~Zapasskii, I.~I.~Ryzhov}








\title{Optical spin noise spectroscopy: application for study 
of gyrotropy spatial correlations}
\maketitle
\begin{abstract}
The single scattering theory is applied to spin noise spectroscopy (SNS). The case of two Gaussian probe beams  tilted with respect to each other is analysed. It is shown that SNS signal in this case carry information about spatial correlations of studied gyrotropic medium. 
\end{abstract}

\section*{Introduction}
Optical spin noise spectroscopy (SNS) first suggested in \cite{Zap} nowadays has demonstrated itself as a powerful method for studying various spin-related phenomena \cite{Zap1,Oest2,Sin}.   By means of SNS it  appeared to be possible to detect magnetic susceptibility of semiconductor systems \cite{Oest1}, to observe  nuclear spin dynamics of these systems\cite{R,R1}, to separate  homogeneous and nonhomogeneous broadened optical lines in vapor\cite{Zap2} etc. 

Despite the fact that even in 1983 it was pointed out that SNS, in fact, is the case of Raman scattering spectroscopy  \cite{Gorb},   the consistent consideration of this effect in terms of single  scattering theory is still of some interest\cite{Scal}.  In this note we present such consideration for Gaussian probe beam and show that slight modification of experimental setup  (adding of additional tilted  probe beam) allows  one to obtaine information concerning the total  space–time correlation function of gyrotropy of studied  system (remind that in conventional SNS only  spatially averaged temporal correlation function is observed).     

The paper is organized as follows.  
  In the first section we present brief explanation of what we call   Gaussian beam  and introduce the  model of polarimetric detector  used in further analysis. 
  In the second section we present the base of single scattering theory and apply it to the medium with randomly distributed in space gyrotropy and calculate  observed polarimetric signal. In  section 3 and 4 we calculate the signal produced by an additional beam (AB) tilted with respect to the main probe beam (AB is out of  detector aperture contrary to  scattered field produced by this beam). We show that this signal is proportional to the Fourier component of gyrotropy spatial correlation function calculated for   wave  vector equal to    difference between  the wave vectors of main and tilted beams. 
In section 5 we present calculations for the model of independent paramagnetic centers and show that  signal produced by an additional tilted probe has the same order of magnitude as the signal produced by  the main probe and, hence,  can be observed using the same experimental setup.        

\section{Gaussian  beam and  the model of polarimetric detector }

In  the simplest consideration the probe beam can be taken in the form of infinite plane wave, but in SNS experiments with two beams described below  this approximation may be not  sufficient  because, as we will see,  the signal produced by  additional probe is proportional to its overlapping with the main probe. For this reason it is convenient to work with Gaussian beams whose   electric  field ${\bf E_p(r)}$ define by the following expression
    \begin{equation}
               {\bf E_p(r)}=e^{\imath (kZ-\omega t)}  
               kQ\sqrt{8 W\over c}
               {(\cos\eta,0,-\sin\eta)\over (2k+\imath Q^2Z)}
               \exp\bigg [-{kQ^2(X^2+y^2)\over 2(2k+\imath Q^2Z)}\bigg ]
               \label{a0}
               \end{equation}
               where ${\bf r}=(x,y,z)$, $k\equiv \omega/c$ ($\omega$ is optical frequency, $c$ is speed of light), $W$ -- beam intensity and 
                           $$
                           \left(\matrix{X\cr Z}\right )=\left(\matrix{\cos\eta &-\sin\eta \cr \sin\eta & \cos\eta}\right )\left(\matrix{x\cr z}\right )
                           $$
    Field (\ref{a0})  satisfy Maxwell's equations and represent the  beam propagating  in $zx$ plane at angle $\eta$ with respect to $z$ axis ($\eta$ is not too large) and polarized mostly in $x$ -direction.  The parameter $Q$ define the $e$-level half width $2w$ of the beam waist  by relationship $w=1/Q$.  $w$ should be greater than  the wavelength $\lambda= 2\pi c/\omega$.  For estimations we accept $\lambda\sim 1\mu$m and $w\sim 30\mu$m.

In SNS experiments small fluctuations of optical field polarization are studied, so, in order to calculate  SNS signal, 
we must specify the model of  polarimetric detector.  We will suppose  the polarimetric detector  to be  comprised of two photodiodes PD1 and PD2 (Figure.\ref{fig1}) placed after polarization beam splitter BS.
 The  output signal  $U$ is obtained by subtraction of diodes photocurrents and  (up to some unimportant coefficient)  can be defined by the expression
\begin{equation}
  U={\omega\over2\pi}\int _{0}^{2\pi/\omega} dt\int_{-l_x}^{l_x}dx\int_{-l_y}^{l_y}dy\bigg [
  \hbox{Re }^2E_x(x,y,L)-\hbox{Re }^2E_y(x,y,L)
   \bigg ],
   \label{a1}
  \end{equation}
where $E_{x,y}$ are $x$ and $y$ projections of  complex input optical field $\bf E$, $2l_{x,y}$ are sizes of sensitive surfaces of  photodiodes along $x$ and $y$ directions.  We ascribe  physical sense to  real part of complex optical field and,  as it is seen from (\ref{a1}), the output  signal $U$ represent the difference between intensities of  input optical field in $x$ and $y$ polarizations integrated over sensitive surfaces of  photodiodes and averaged over one  optical period $2\pi/\omega$.  
   
In our case the input optical field $\bf E$ can be presented as a sum of   the probe field  $\bf E_0$ (Re $\bf E_0={\cal E}_0$) and the field $\bf E_1$ (Re $\bf E_1={\cal E}_1$)  aroused  due to    scattering of the probe  beam by   a sample with randomly distributed in space gyrotropy. Then, the first order with respect to $\bf E_1$ contribution $u_1$ to   polarimetric signal  can be written as
 \begin{equation}
 u_1={\omega\over\pi}\int _{0}^{2\pi/\omega} dt\int_{-l_x}^{l_x}dx\int_{-l_y}^{l_y}dy 
\bigg [
{\cal E}_{x0}(x,y,L){\cal E}_{x1}(x,y,L)-{\cal E}_{y0}(x,y,L){\cal E}_{y1}(x,y,L)\bigg ]
\label {a2} 
\end{equation}

In the next section we develop the theory for calculating scattered field ${\bf{\cal E}} _1$ .

\section{Single scattering theory for the medium with random gyrotropy}

In this section we consider scattering of monochromatic beam by a medium with random gyrotropy. It means that medium polarization  ${\bf P(r) }$    related to electric field  ${\bf E(r)}$ by:
\begin{equation}
{\bf P(r)}=\imath [{\bf E(r)G(r)}]=\imath {\bf E(r)\times G(r)}
\label{a3}
\end{equation}
where ${\bf G(r)}$ is space dependent gyration vector. At this stage of our treatment we suppose the gyration vector to be time independent. Then,   Maxwell's equations for  the electromagnetic field  in considered medium can be reduced to the following equation
\begin{equation}
\Delta {\bf E}+k^2{\bf E}=-4\pi k^2{\bf P}-4\pi \hbox{ grad div }{\bf P}, \hskip10mm k\equiv {\omega\over c}
\label{a4}
\end{equation}
We will search for the solution of this equation in the form of series in powers of ${\bf G(r)}$. The zero order term ${\bf E_0(r)}$  represents the probe beam field  which we consider to be known. The first order term ${\bf E_1(r)}$ corresponds to  single scattering approximation which is sufficient for our purposes.  
This term  satisfies  the equation
\begin{equation}
\Delta {\bf E_1}+k^2{\bf E_1}=-4\pi\imath k^2{\bf E_0(r)\times G(r)}-4\pi\imath \hbox{ grad div }{\bf E_0(r)\times G(r)}
\label{a5}
\end{equation}

The solution of this equation can be expressed in terms of  Green’s 
function $\Gamma({\bf r})=-\cos (kr)/4\pi r$ of Helmholtz equation $[\Delta+k^2]\Gamma ({\bf r})=\delta({\bf r})$:
\begin{equation}
     {\bf E_1(r)}=\imath\int  {\cos (k|{\bf r-r'}|)\over {\bf |r-r'|}}\bigg [ k^2{\bf E_0(r')\times G(r')}+ \hbox{ grad div }{\bf E_0(r')\times G(r')}\bigg ]d^3{\bf r'}
      \label{a6}
      \end{equation}
 Let the sample (we call  “sample”  the region where ${\bf G(r)}$ is  nonzero)  be placed in the vicinity of the origin of our coordinate system $x,y,z$.  Let   the photosensitive surface   of polarimetric detector be parallel to $xy$ plane and detector itself be placed  at $z=L$ with $L$ large  as compared with the sample sizes.  Then,  as it is seen from (\ref{a6}) ,  the scattered field can be written as a sum  of two contributions:   
\begin{equation}
{\bf E_1(r)=E_{1}^1(r)+E_{1}^2(r)}
\label{a7}
\end{equation}
$$
{\bf E_{1}^1(r)}\equiv {\imath k^2\over L}\int  \cos (k|{\bf r-r'}|){\bf E_0(r')\times G(r')} d^3{\bf r'}
$$
$$
{\bf E_{1}^2(r)}\equiv {\imath\over L}\int  \cos (k|{\bf r-r'}|)\hbox{ grad div }{\bf E_0(r')\times G(r')}d^3{\bf r'}={1\over k^2}\hbox{ grad div }{\bf E_1^1(r)}
$$

We will concentrate our efforts on calculating  the part ${\bf E_1^1(r)}$  of scattered field  because      below we will  need this  field  for small scattering angles  and in this case, as it can be directly checked, only ${\bf E_1^1(r)}$ is important.           
 We will take the probe beam in the form (\ref{a0}) at $\eta =0$ rotated by angle $\phi$ around $z$-axis for this angle to specify the beam polarization in $xy$ plane.  
Hence,  the probe beam field has the form 
  \begin{equation}
  {\bf E_0(r)}=e^{\imath [kz-\omega t]}  
   kQ\sqrt{8 W\over c} {(\cos\phi,\sin\phi,0)\over (2k+\imath Q^2z)}
      \exp\bigg [-{kQ^2(x^2+y^2)\over 2(2k+\imath Q^2z)}\bigg ], \hskip10mm {\bf r}=(x,y,z)
                         \label{a8}
                         \end{equation}
We need  this field in two considerably  separated spatial regions:   firstly,  in Eq.(\ref{a2}) at large  values of $z\sim L$ and,     secondly, in Eq.(\ref{a7}) at relatively small values of $z$  within the sample. Calculation for $z\sim L$  show that  field ${\bf {\cal E}}_0$ entering Eq.(\ref{a2}) has the form
     
 \begin{equation}
               \left (\matrix {{\cal E}_{x0}(x,y,L)\cr {\cal E}_{y0}(x,y,L)}\right )=
\left (\matrix {\cos\phi\cr\sin\phi}\right ) \sqrt{8 W\over c}
             {k\over  QL}\hskip1mm \sin \bigg [ kL-\omega t
             +{k[x^2+y^2]\over 2L}
             \bigg ]  
                                               \exp\bigg [-{k^2(x^2+y^2)\over  Q^2L^2}\bigg ]
                                               \label{a9}
                                               \end{equation}

While obtaining these expressions we accept that $L> z_c =4\pi w^2/\lambda$ ($z_c$  is
 Rayleigh length). To calculate the scattered field by Eq. (\ref{a7}) one need the 
field (\ref{a8}) at $z<z_c$. In this limit Eq. (\ref{a8}) can be 
simplified as follows

\begin{equation}
                         {\bf E_0(r)}=e^{\imath [kz-\omega t]}  
                         Q\sqrt{8 W\over c}
                         {(\cos\phi,\sin\phi,0)\over 2}
                         \exp\bigg [-{Q^2(x^2+y^2)\over 4}\bigg ],\hskip10mm z<z_c
                         \label{a10}
                         \end{equation}
Using this relationship one can calculate the scattered field 
${\bf E_1^1(r)}$ (\ref{a7}) and obtain for  real parts ${\cal E}_{x1}$ and ${\cal E}_{y1}$  
entering the Eq. (\ref{a2}) the following expressions
 
 \begin{equation}
\left (\matrix{{\cal E}_{x1} \cr
{\cal E}_{y1}}\right )
=\left (\matrix{-\sin\phi\cr\cos\phi}\right ) \sqrt{2 W\over c}{Q k^2\over L}
 \int  \cos (k|{\bf r-r'}|) 
\sin [kz'-\omega t]                           
                         \exp\bigg [-{Q^2(x'^2+y'^2)\over 4}\bigg ]
 G_z({\bf r'}) d^3{\bf r'}
\label{a11}
\end{equation}

Using   the explicit expressions (\ref{a9}) and (\ref{a11}) for  probe ${\cal E}_0$ and scattered ${\cal E}_1$ field 
we can calculate  polarimetric signal by  Eq.  (\ref{a2}). While averaging  the product 
 ${\cal E}_{x0}{\cal E}_{x1}$ over one optical period the following integral is encountered
    $$
  {\omega\over \pi}\int_0^{2\pi/\omega}{\cal E}_{0x}{\cal E}_{1x}\hskip1mm dt\sim
 {\omega\over \pi}\int_0^{2\pi/\omega}\sin [kz'-\omega t]\hskip1mm
 \sin \bigg [ kL-\omega t
              +{k[x^2+y^2]\over 2L}
              \bigg ]  \hskip1mm dt =
  $$
$$
=\cos \hskip1mm k\bigg [
z'-L-{x^2+y^2\over 2L}
\bigg ]
$$
  The same for  ${\cal E}_{y0}{\cal E}_{y1}$. Then,  Eq. (\ref{a2}) gives

 \begin{equation}
    u_1= -{4Wk^3 \sin[2\phi]\over cL^2}                     
                     \int_{-l_x}^{l_x}dx\int_{-l_y}^{l_y}dy 
                                          \exp\bigg [-{k^2(x^2+y^2)\over  Q^2L^2}\bigg ]
                                          \times
           \label {a12} 
\end{equation}
$$
\times \int  \cos (k|{\bf r-r'}|) 
\hskip1mm
\cos \hskip1mm k\bigg [
z'-L-{x^2+y^2\over 2L}
\bigg ]                         
      \hskip1mm \exp\bigg [-{Q^2(x'^2+y'^2)\over 4}\bigg ]
          G_z({\bf r'}) d^3{\bf r'},
$$
with  ${\bf r}=(x,y,L)$ and ${\bf r'}=(x',y',z')$. The external integration over $dxdy$ runs over the  detector
 surface, therefore $|x|,|y|<l_{x,y}\ll L$. We will suppose that  detector sizes $l_{x,y}$  
 exceed the size $L\lambda/2\pi w$ of the probe beam spot at the detector 
  (see Eq. (\ref{a9})).  Then, $x$ and $y$ can be eatimated as $x,y\sim L\lambda/2\pi w$.
  The internal integration $d{\bf r'}$ runs over the 
 irradiated volume of the sample. For this reason $x',y'\sim w$ and $z'$ is of the order of sample length $l_s$.
 Taking into account that $L\lambda/2\pi w,w,l_s\ll L$ we can obtaine 
  the following expansion for the factor $|{\bf r-r'}|$:
 
   \begin{equation}
   |{\bf r-r'}|\approx L+{x^2+y^2\over 2L}+
    {x'^2+y'^2\over 2L}-
   {xx'+yy'\over L}-{z'}.
  \label{a13}
    \end{equation}
Note that term $\sim z'^2$ vanishes. Further estimations show that  term $(x'^2+y'^2)/2L$
 can be omitted because in our case $k(x'^2+y'^2)/2L<\pi/4$ and finally we have
 \begin{equation}
   |{\bf r-r'}|\approx
 L+{x^2+y^2\over 2L}
    -{z'}
   -{xx'+yy'\over L}
  \label{a14}
   \end{equation} 
 
Using this formula one can evaluate  the product of cosine functions in (\ref{a12}) as

   \begin{equation}
   \cos (k|{\bf r-r'}|) \hskip1mm \cos \hskip1mm k
   \bigg [
   z'-L-{x^2+y^2\over 2L}
   \bigg ]  =  
   \label{a15}
   \end{equation}
   $$
   =\cos  k\bigg [
       z'-L-{x^2+y^2\over 2L}
         +{xx'+yy'\over L}
      \bigg ]
      \hskip1mm \cos \hskip1mm k
         \bigg [
         z'-L-{x^2+y^2\over 2L}
         \bigg ]=           
   $$
   $$
  = {1\over 2}\bigg \{
   \cos  k\bigg [
          {xx'+yy'\over L}
         \bigg ]+
    \cos  k\bigg [
           2(z'-L)-{x^2+y^2\over L}
             +{xx'+yy'\over L}
          \bigg ]
   \bigg \}
   $$
 
 As it was mentioned above, the detector sizes  considered to be  greater than the probe beam spot: 
$l_{x,y}> L\lambda/2\pi w$.
 This allows one to extend  integration over  the detector surface in (\ref{a12}) to infinity: $|l_{x,y}|\rightarrow\infty$
 and calculate all  integrals using the   following formula   
      \begin{equation}
       \int dx \exp [-\alpha x^2+\imath\beta x ]=\sqrt{\pi\over \alpha}\exp \bigg (-{\beta^2\over 4\alpha}\bigg ).
       \label{a16}
       \end{equation}
  For example, the integral with the first cosine function in  Eq.(\ref{a15}) (we denote it $I_1$) can be calculated as follows:

   \begin{equation}
   I_1=\int_{-\infty}^\infty dx \int_{-\infty}^\infty dy\hskip1mm  \exp\bigg [-{k^2(x^2+y^2)\over  Q^2L^2}\bigg ] \hskip1mm \cos  k\bigg [
             {xx'+yy'\over L}
            \bigg ]=
\label{a17}   
\end{equation}
   $$
   =\hbox{ Re }
   \int_{-\infty}^\infty dx \int_{-\infty}^\infty dy\hskip1mm  \exp\bigg [-{k^2(x^2+y^2)\over  Q^2L^2}+\imath k 
                   {xx'+yy'\over L}
                     \bigg ]= 
  $$             
   $$
      =\hbox{ Re }
      \int_{-\infty}^\infty dx 
       \exp\bigg [-{k^2 x^2\over  Q^2L^2}+\imath k 
                            {xx'\over L}
                              \bigg ]
            \hskip2mm\int_{-\infty}^\infty dy\hskip1mm  \exp\bigg [-{k^2y^2\over  Q^2L^2}+\imath k 
                      {yy'\over L}
                     \bigg ]=
     $$  
     $$
     ={\pi Q^2L^2\over k^2} \exp  \bigg (
                -{[x'^2+y'^2]Q^2\over 4}
                \bigg )  
         $$                  
      The integral with  second cosine function in Eq. (\ref{a15}) calculated in the same manner
 and  can be neglected as compared with (\ref{a17})  
 because $Q^2L=L/w^2\gg k=2\pi/\lambda$.
 
Substituting (\ref{a17}) to (\ref{a12}) we obtain the following expression for   polarimetric signal:

            \begin{equation}
                  u_1= -{4Wk \pi Q^2\sin[2\phi]\over c}                     
               \int_V   
                   \exp\bigg [-{Q^2(x'^2+y'^2)\over 2}\bigg ]
                            G_z({\bf r'}) d^3{\bf r'}                   
                         \label {a18} 
              \end{equation}

Remind that this formula is valid if  sample length $l_s$ is smaller than Rayleigh length $l_s<z_c$ (see definition of Rayleigh length after Eq. (\ref{a9})) and
 the probe beam spot is less than detector photo sensitive surface $l_{x,y}\gg L\lambda/2\pi w$. It is
 seen from Eq. (\ref{a18}) that, in fact,   polarimetric signal is proportional to $z$-component of gyration
 averaged over  irradiated volume of the sample as it is usually supposed intuitively.

Eq. (\ref{a18}) allows  one to obtain the expression for power spectrum of magnetization noise observed in SNS.  In this case ${\bf G(r)}$ is  proportional to spontaneous magnetization of the sample which represents  space-time    random field. If characteristic  frequencies of this field are much lower than optical frequency $\omega$  one can use Eq. (\ref{a18}) for calculation of random polarimetric signal by replacing ${\bf G(r)}\rightarrow {\bf G(r,}t)$.
 Noise power spectrum ${\cal N}(\nu)$ is defined as Fourier transformation of  the correlation function of polarimetric signal. 
 Using Eq. (\ref{a18}) the noise power spectrum ${\cal N}(\nu)$ can be expressed in terms  of the  gyrotropy ${\bf G(r,}t)$ space-time correlation function :
\begin{equation}
{\cal N}(\nu)=\int dt\langle u_1(t)u_1(0)\rangle e^{\imath \nu t}={16 W^2k^2\pi^2Q^4\sin^2[2\phi]\over c^2}\times
\label{a19}
\end{equation}
$$
\times
\int dt\hskip1mm  e^{\imath \nu t}\int_V d^3{\bf r}\int_V d^3{\bf r'}
\exp\bigg [-{Q^2(x'^2+y'^2+x^2+y^2)\over 2}\bigg ]
                                     \langle G_z({\bf r'},0)G_z({\bf r},t)\rangle
  $$

\begin{figure}
    \includegraphics[width=.8\columnwidth,clip]{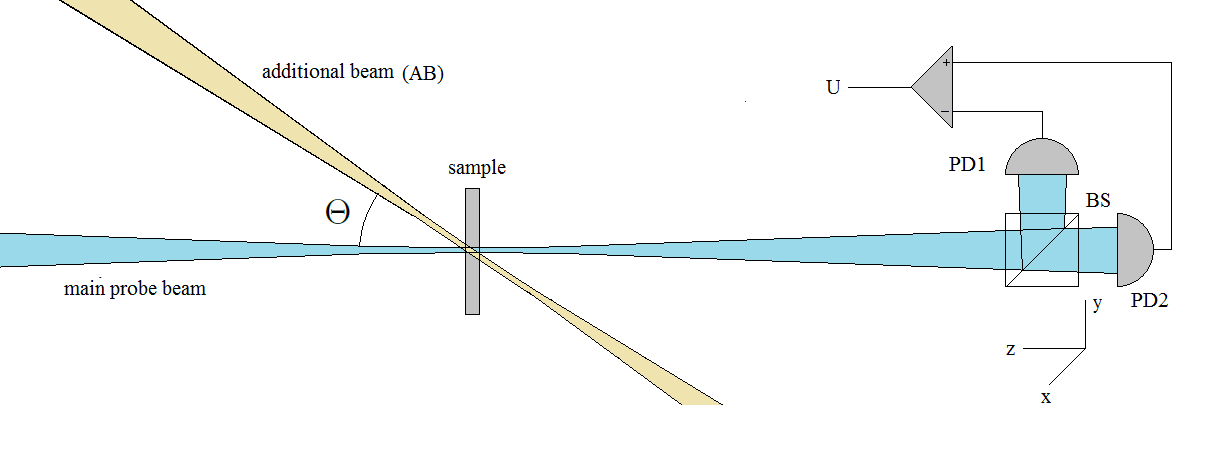}
     \caption{Registration    of noise signal produced by additional tilted beam }
     \label{fig1}
     \end{figure}

  To calculate the correlation function $ \langle G_z({\bf r'},0)G_z({\bf r},t)\rangle$ entering (\ref{a19}) one should specify a concrete model of gyratropic medium. The example of such model (the   model  of independent  paramagnetic atoms with fluctuating magnetization) will be described in section 5.      In the next section we will calculate  the plarimetric signal produced by additional tilted beam which give rise to scattered field but do not irradiate  the detector (see Figure \ref{fig1}). 
   
\section{Additional tilted beam }

Let us irradiate our sample by an additional beam (AB) tilted by angle $\Theta$ with respect to the main probe beam  (Figure (\ref{fig1})).  Note that AB do not hit the detector, but   scattered field produced by AB  can give rise to additional polarimetric signal and our goal now  is to calculate it.    
 The calculation can be performed in the same manner as in the previous section  with  following changes. 
The scattered field can be calculated by Eq. (\ref{a7})  in which the field ${\bf E_0(r)}$ should be replaced by $ {\bf E_0^t(r)}$, where $ {\bf E_0^t(r)}$ represents the field of tilted  beam. The field $ {\bf E_0^t(r)}$ can be obtained by  rotation of  ${\bf E_0(r)}$ by angle $\Theta$ around the axis $(\cos\phi,\sin\phi,0)$ parallel to the  direction of polarization of the probe beam \footnote{For this reason  polarization of tilted  and  probe beams is the same}:
\begin{equation}
 {\bf E_0^t(r)}=M {\bf E}_0(M{\bf r}).
 \label{a20}
  \end{equation}
Here matrix $M$ is defined as

 \begin{equation}
  M=R(-\phi)H(\Theta)R(\phi)=\left (\matrix {\cos\Theta\sin^2\phi +\cos^2\phi&[1-\cos\Theta]\sin \phi\cos\phi & -\sin\phi\sin\Theta \cr [1-\cos\Theta]\sin\phi\cos\phi& \cos\Theta\cos^2\phi+\sin^2\phi &\cos\phi\sin\Theta \cr \sin\Theta\sin\phi &-\sin\Theta\cos\phi & \cos\Theta}\right )=
  \label{a21}
  \end{equation}
  $$
  =\left (\matrix {1-{1\over 2}\Theta^2\sin^2\phi &{1\over 2}\Theta^2\sin \phi\cos\phi & -\Theta\sin\phi \cr {1\over 2}\Theta^2\sin\phi\cos\phi & 1- {1\over 2}\Theta^2\cos^2\phi &\Theta\cos\phi \cr \Theta\sin\phi &-\Theta\cos\phi & 1-{1\over 2}\Theta^2}\right )+O(\Theta^3)
  $$
 Therefore, ${\bf E_0^t(r)}$ defined by the following expression
 \begin{equation}
                           {\bf E_0^t(r)}=  
          Q\sqrt{2 W_t\over c}
                           (\cos\phi,\sin\phi,0)\hskip1mm 
     \exp\imath\bigg [k Z({\bf r})-\omega t\bigg ]                                                 
                           \hskip1mm \exp\bigg [-{Q^2[X^2({\bf r})+Y^2({\bf r})]\over 4}\bigg ]
                           \label{a22}
                           \end{equation}
where 

 \begin{equation}
  \left (\matrix {X ({\bf r})\cr Y({\bf r})\cr Z({\bf r})}\right )\equiv
  \left (\matrix {\cos\Theta\sin^2\phi +\cos^2\phi&[1-\cos\Theta]\sin \phi\cos\phi & -\sin\phi\sin\Theta \cr [1-\cos\Theta]\sin\phi\cos\phi& \cos\Theta\cos^2\phi+\sin^2\phi &\cos\phi\sin\Theta \cr \sin\Theta\sin\phi &-\sin\Theta\cos\phi & \cos\Theta}\right )\left (\matrix {x \cr y\cr z}\right ) +\left (\matrix {\delta x \cr \delta y \cr \delta z }\right )
  \label{a23}
  \end{equation}
with ${\bf r}=(x,y,z)$. We denote by $W_t$ the intensity of AB and take into account the possible spatial shift $(\delta x,\delta y,\delta z)$ of AB. 
 Substituting ${\bf E_0^t(r’)}$  (\ref{a22})  to Eq. (\ref{a7}) instead ${\bf E_0(r’)}$, one can obtain   the following expression for scattered field produced by AB:
                                     
   \begin{equation}
      \left (\matrix {{\cal E}_{1x}^t\cr {\cal E}_{1y}^t}\right ) =   
      \left (\matrix {-\sin\phi \cr \cos\phi }\right )
      \sqrt{2 W_t\over c}{Q k^2\over L}
       \int  \cos (k|{\bf r-r'}|) 
      \sin [kZ'-\omega t]                           
                               \exp\bigg [-{Q^2(X'^2+Y'^2)\over 4}\bigg ]
       G_z({\bf r'}) d^3{\bf r'}
      \label{a24}
      \end{equation}
      where $X'=X({\bf r'})$, $Y'=Y({\bf r'})$ and $Z'=Z({\bf r'})$, with functions           $X({\bf r'})$,                       
     $Y({\bf r'})$, 
     $Z({\bf r'})$  defined by Eq. (\ref{a23}), in which $x,y,z\rightarrow x',y',z'$.
      This formula has the same sense as Eq. (\ref{a11}); for clarity we supply the components of scattered field by $t$.
      Taking into account this replacements, one can get the relationship for polarimetric signal produced by AB (instead of Eq. (\ref{a12}))
                                       
       \begin{equation}
           u_1^t= -{4\sqrt{WW_t}k^3 \sin[2\phi]\over cL^2}                     
                            \int_{-l_x}^{l_x}dx\int_{-l_y}^{l_y}dy 
                                                 \exp\bigg [-{k^2(x^2+y^2)\over  Q^2L^2}\bigg ]
                                                 \times
                  \label {a25} 
       \end{equation}
       $$
       \times \int  \cos (k|{\bf r-r'}|) 
       \hskip1mm
       \cos \hskip1mm k\bigg [
       Z'-L-{x^2+y^2\over 2L}
       \bigg ]                         
             \hskip1mm \exp\bigg [-{Q^2(X'^2+Y'^2)\over 4}\bigg ]
                 G_z({\bf r'}) d^3{\bf r'}
       $$ 
Calculation of intergrals  can be carried out in the same manner as it was done in the previous section and  the final result for the polarimetric signal produced by AB is:   
  \begin{equation}
                      u_1^t= -{4\sqrt{WW_t}k \pi Q^2\sin[2\phi]\over c}                     
                   \int_V   \cos k[z'-Z']
                       \exp\bigg [-{Q^2(x'^2+y'^2+X'^2+Y'^2)\over 4}\bigg ]
                                G_z({\bf r'}) d^3{\bf r'}                   
                             \label {a26}  
                 \end{equation}           
                 where                        ${\bf r'}=(x',  y',  z')$
          \begin{equation}
          \left (\matrix {X' \cr Y' \cr Z'}\right )=
          \left (\matrix {\cos\Theta\sin^2\phi +\cos^2\phi&[1-\cos\Theta]\sin \phi\cos\phi & -\sin\phi\sin\Theta \cr [1-\cos\Theta]\sin\phi\cos\phi& \cos\Theta\cos^2\phi+\sin^2\phi &\cos\phi\sin\Theta \cr \sin\Theta\sin\phi &-\sin\Theta\cos\phi & \cos\Theta}\right )\left (\matrix {x' \cr y' \cr z'}\right ) +\left (\matrix {\delta x \cr \delta y \cr \delta z }\right )
          \label{a27}
          \end{equation}   
          Note that  $u_1^t$ is proportional to overlapping between the probe beam and AB and vanish at sufficiently large shift $\delta x,\delta y,\delta z$.   Trigonometric factor $\cos k[z’-Z’]$ is, in fact, emphasize the  harmonic of gyrotropy with spatial frequency equal to  difference between the wave vectors of  probe beam and AB.                    
          Total signal in presence of both probe and AB is the sum of  (\ref{a18})  and (\ref{a26}): $u_1+u_1^t$.   Remind that angle $\Theta$  should be  not too large. In the opposite case one should take into account the component ${\bf E_1^2(r)}$ in Eq. (\ref{a7}).

        \section{ Noise signal produced by probe and additional beams simultaneously  } 
           
           Noise signal produced   by   probe beam and AB is calculated as Fourier transform of  the total polarimetric signal  $u=u_1+u^t_1$  correlation function. It  consists of 3 terms: 
\begin{equation}
{\cal N}_t(\nu)= \int dt e^{\imath \nu t}\langle u(0)u(t)\rangle=\int dt e^{\imath \nu t}\bigg [\langle u_1(0)u_1(t)\rangle+2\langle u_1(0)u_1^t(t)\rangle+\langle u_1^t(0)u_1^t(t)\rangle\bigg ]
\label{a28}
 \end{equation}
Using Eq's (\ref{a18})  and (\ref{a26}) one can write the  expressions for each of these terms. The first term has been already calculated and is difined by Eq. (\ref{a19}). For  correlator entering the last term we have the following expression  

   \begin{equation}
      \langle u_1^t(0)u_1^t(t)\rangle=
      {16 WW_tk^2\pi^2Q^4\sin^2[2\phi]\over c^2}     
            \int_V d^3{\bf r}\int_V d^3{\bf r'}
            \cos k[z-Z]\cos k[z'-Z']\times
      \label{a29}
      \end{equation}
      $$
      \times\exp\bigg [-{Q^2(X^2+Y^2+x^2+y^2+X'^2+Y'^2+x'^2+y'^2)\over 4}\bigg ]\times
                                               \langle G_z({\bf r'},0)G_z({\bf r},t)\rangle,
 $$         
       
    where  $x,y,z\rightarrow {\bf r}$  and $X,Y,Z$ are defined by relationship Eq. (\ref{a27})  
    
            \begin{equation}
            \left (\matrix {X \cr Y \cr Z}\right )=
            \left (\matrix {\cos\Theta\sin^2\phi +\cos^2\phi&[1-\cos\Theta]\sin \phi\cos\phi & -\sin\phi\sin\Theta \cr [1-\cos\Theta]\sin\phi\cos\phi& \cos\Theta\cos^2\phi+\sin^2\phi &\cos\phi\sin\Theta \cr \sin\Theta\sin\phi &-\sin\Theta\cos\phi & \cos\Theta}\right )\left (\matrix {x \cr y \cr z}\right ) +\left (\matrix {\delta x \cr \delta y \cr \delta z }\right )
            \label{a30}
            \end{equation} 
         $X',Y',Z'$ are  similar  functions of  $x',y',z'\rightarrow {\bf r'}$.
      
      Finally, the cross correlator $\langle u_1^t(0)u_1(t)\rangle$ can be written as
     \begin{equation}
               \langle u_1^t(0)u_1(t)\rangle=
               {16 W\sqrt{WW_t}k^2\pi^2Q^4\sin^2[2\phi]\over c^2}     
                     \int_V d^3{\bf r}\int_V d^3{\bf r'}
                     \cos k[z-Z]\times
               \label{a31}
               \end{equation}
               $$
               \times\exp\bigg [-{Q^2(X^2+Y^2+x^2+y^2)\over 4}-{Q^2(x'^2+y'^2)\over 2}\bigg ]\times
                                                        \langle G_z({\bf r'},0)G_z({\bf r},t)\rangle
          $$
    Let us now consider the sence of various  factors entering     Eqs.(\ref{a19},\ref{a29},\ref{a31}).

{\bf Exponential factor }
  reduce the region of integration to the region of overlapping of  the probe beam and AB. If $\Theta$ is not too large and $l_s\Theta < w$, this region is close to “beam volume within the sample”.  In this case exponential factor can be calculated at
 $X=x,Y=y,Z=z,X'=x',Y'=y',Z'=z'$. Note that  it is rather difficult  to satisfy the condition $ l_s \Theta < w$   in real experiment. For this reason   overlapping factor may  considerably reduce the contribution of AB to polarimetric signal.     
    
 {\bf Trigonometric factor } at small angles $\Theta$ depends on  difference between  wave vectors of probe beam and AB because  cosine argument can be evaluated  as     $z-Z=[\cos\phi y-\sin\phi x]\Theta$. 
  
  {\bf Correlation function } $\langle G_z({\bf r'},0)G_z({\bf r},t)\rangle$ defined by a 
concrete model of gyrotropic medium. For homogeneous mediums it depends on  the difference ${\bf r-r'}$ 
of space arguments. For the model of independent paramagnetic centers described below
$\langle G_z({\bf r'},0)G_z({\bf r},t)\rangle$ $\sim \delta ({\bf r-r'})e^{-|t|/\tau}\cos\omega_0t$

If   concrete model of gyrotropic medium is specified, then the integrals entering Eqs.(\ref{a19},\ref{a29},\ref{a31}) can
 be calculated in the frame of this model. In the next section  we will present calculation for  the model 
of independent paramagnetic centers. Nevertheless, the following general remarks can be made.
Suppose that  beam waste $4w$ and   sample length $l_s$ are much greater than   
correlation radius of  gyrotropy  $R_c$ and  spatial period $2\pi/k\Theta$  related to 
 the difference of wave vectors of the probe beam and AB: $4w,l_s\gg R_c,2\pi/k\Theta$.
 Then, one can replace variables in integrals entering Eqs.(\ref{a19},\ref{a29},\ref{a31}) 
in the following way: ${\bf r,r'}\rightarrow {\bf R\equiv  r-r',R'\equiv r+r'}$ 
 and take advantage of the fact that correlator 
  $ \langle G_z({\bf r'},0)G_z({\bf r},t)\rangle$
depends on  difference of its arguments:
\begin{equation}
  \langle G_z({\bf r'},0)G_z({\bf r},t)\rangle\equiv K({\bf r-r'},t)
\label{a32}
\end{equation}

Then,  the integral over  ${\bf R\equiv r-r'}$  in Eq. (\ref{a19}) can be estimated as  
the average of $K({\bf R},t)$ over  irradiated volume of the sample $V_b $. The integration over
 ${\bf R'\equiv r+r'}$ gives this  volume itself and we obtain    

\begin{equation}
{\cal N}(\nu)={16 W^2k^2\pi^2Q^4\sin^2[2\phi]\over c^2}
\int dt\hskip1mm  e^{\imath \nu t}\int_V d{\bf r} d{\bf r'}
\exp\bigg [-{Q^2(x'^2+y'^2+x^2+y^2)\over 2}\bigg ]K({\bf r-r'},t)\sim
\label{a33}
\end{equation}
$$
\sim {W^2l_s\sin^2[2\phi]\over S}\int dt\hskip1mm e^{\imath\nu t}\int_{V_b} d{\bf R} \hskip1mm K({\bf R},t).
  $$                  
        Here we denote the cross section area of the beam by $S\equiv 4\pi w^2$ and take into account that $w=1/Q$ and 
 that irradiated  volume of the sample is 
 $V_b= Sl_s$, where $l_s$
 is the sample length.

The correlation function Eq.(\ref{a29}) can be estimated in a similar way.  
If $\Theta$ is not too large the arguments of cosine functions can be evaluated as:   
$z-Z=[\cos\phi y-\sin\phi x]\Theta$ and $z'-Z'=[\cos\phi y'-\sin\phi x']\Theta$.
  Therefore,  one can represent the  product of cosine functions in  Eq.(\ref{a29}) as
    $$
    \cos k[z-Z]\cos k[z'-Z']={1\over 2}\cos \bigg\{k\Theta \bigg [(y-y')\cos\phi -(x- x')\sin\phi \bigg ] \bigg\} +
    $$
$$
+{1\over 2}\cos \bigg\{k\Theta \bigg [(y+y')\cos\phi -(x+x')\sin\phi \bigg ] \bigg\}
$$

Note that  difference $\Delta {\bf k}$ between the wave vector of the probe 
beam and AB for small $\Theta$ has only $x$ and $y$ components: $\Delta {\bf k}=k\Theta (-\sin\phi,\cos\phi,0)$.
Therefore,  this relationship after replacement of variables  
${\bf r,r'}\rightarrow {\bf R= r-r',R'=r+r'}$ takes the form
    $$
    \cos k[z-Z]\cos k[z'-Z']={1\over 2}\cos (\Delta {\bf k,R}) +
{1\over 2}\cos (\Delta {\bf k,R'})
    $$

Remind that our treatment is valid if
 $w$ is large enough for $\Delta kw>2\pi$. In this case 
 the integral $\int_{V_b}d{\bf R'}\cos (\Delta {\bf k,R'})\sim 0$
 and we come to the conclusion that  
  correlation function Eq. (\ref{a29}) can be estimated as follows 

   \begin{equation}
      \langle u_1^t(0)u_1^t(t)\rangle\sim
      {WW_t Q^4\sin^2[2\phi]}     
            \int_{V_b} d{\bf R}d{\bf R'}\hskip1mm K({\bf R},t)
\cos (\Delta {\bf k,R})  \sim
      \label{a34}
      \end{equation}
  $$
\sim  {WW_t \sin^2[2\phi]l_s\over S}     
            \int_{V_b} d{\bf R}  \hskip1mm K({\bf R},t) \cos (\Delta {\bf k,R})         
$$

 Therefore, the contribution of tilted AB to  noise signal is proportional to the Fourier 
transform of  correlation function of gyrotropy calculated at differential wave vector 
of the probe and additional beams $\Delta {\bf k}$.
 Analogously  it can be shown that  contribution of  cross correlator 
Eq. (\ref{a31}) is relatively small under considered conditions.

   \section { Model of independent paramagnetic centers }
It this model the random field of gyrotropy $G_z({\bf r})$ has the form
    \begin{equation}
      G_z({\bf r})=\sum_{i=1}^N g_i(t)\delta ({\bf r-r_i}),
      \label{a35}
      \end{equation}
corresponding  to $N$ paramagnetic centers  randomly distributed in space with
  some density $\sigma$ defined as $\sigma\equiv N/V$ where $V$ is total  volume of the system.
  We accept that $g_i(t)$ is proportional to $z$-component of magnetization of $i$-th center.     The polarimetric signal can be calculated by Eq.(\ref{a18}):

 \begin{equation}
                              u_1=u_1(t)= -{4Wk \pi Q^2\sin[2\phi]\over c}                     
                           \int_V   
                               \exp\bigg [-{Q^2(x'^2+y'^2)\over 2}\bigg ]
                                         \sum_i g_i(t)\delta ({\bf r'-r_i})d^3{\bf r'}                   
                                     \label {a36} 
                          \end{equation}
 Let us calculate  polarimetric signal $u_{10}$ produced by  a sample in which all
 magnetizations $g_i(t)$ are constant and equal to each other: $g_i(t)=g_0=$ const.  
This corresponds to  paramagnet  placed in a high magnetic field at low temperatures.  Formula Eq. (\ref{a36}) gives 

      \begin{equation}
  u_{10}= -{8Wk g_0\sigma l_s\pi^2 \sin[2\phi]\over c}                     
          \label {a37} 
    \end{equation}    
 We will see below that $u_{10}$ is a convenient scale. 
 Let us now consider  the state  of our gyrotropic medium  for which   the quantities $g_i$ are not constants but represent stationary random processes with  following correlation function:  $\langle g_i(t)g_k(t')\rangle =\delta_{ik}K(t-t’)$ (do not confuse with space-time correlation function Eq. (\ref{a32})) and calculate the noise power spectrum Eq. (\ref{a19}) for this model. We have
 \begin{equation}
  \langle G_z({\bf r'},0)G_z({\bf r},t)\rangle={1\over V}\sum_{i}\int d^3{\bf r_i}
  \langle g_i(0)g_i(t)\rangle \delta ({\bf r'-r_i}) \delta ({\bf r-r_i})=\delta ({\bf r-r'})\sigma K(t),
  \label{a38}
  \end{equation}     
and, consequently 
    \begin{equation}
          {\cal N}(\nu)={16 W^2k^2\pi^3Q^2l_s\sigma\sin^2[2\phi]\over c^2}     
          \int dt\hskip1mm  e^{\imath \nu t} K(t)                 
\label{a39}          
\end{equation}

If we accept for  beam area the expression $S=4\pi w^2$, then $Q^2=1/w^2=4\pi/S$.  Taking  into account Eq. (\ref{a37})   we obtain the following relationship for noise power spectrum

\begin{equation}
{\cal N}(\nu)={u_{10}^2\over  \sigma l_sS}\hskip1mm \int dt e ^{\imath \nu t}{\langle g(0)g(t)\rangle\over g_0^2}
\label{a40}
\end{equation}     

Note that   $\sigma  l_s S\equiv N_b$ is the number of centers in the irradiated  volume of the sample.

 In the simplest case each  paramagnetic center  of our gyrotropic medium can be associated with effective  spin 1/2 . In this case the total magnetization can be expressed as:  $g_0^2=(g\beta)^2 /4$ (here $g$ is effective  $g$-factor and $\beta$ is Bohr magneton). If the transverse magnetic field $B_x$ is applied    the correlator $\langle g(0)g(t)\rangle$ can be calculated by following chain of relationships:
\begin{equation}
\langle g(0)g(t)\rangle ={(g\beta)^2 \over 2}\hbox{ Sp }[S_zS_z(t)+S_z(t)S_z]\rho_{eq}\hskip 10mm
S_z(t)=e^{-\imath \omega_0tS_x}S_ze^{\imath \omega_0tS_x} \hskip5mm \omega_0\equiv {g\beta B_x\over\hbar}
\label{a41} 
\end{equation}    
Here $\rho_{eq}$ is the density matrix of  two-level system representing our effective spin 1/2.
 If temperature is high enough for $kT\gg g\beta B_x$, the density matrix can be taken constant
 $\rho_{eq}=\hat I/2$ ($\hat I$  is the unit matrix)  and we obtain 
\begin{equation}
\langle g(0)g(t)\rangle = {(g\beta)^2\over 4}\hbox{ Sp }[S_zS_z(t)+S_z(t)S_z] 
={(g\beta)^2\over 4}\hskip1mm \cos\omega_0 t \rightarrow 
{(g\beta)^2\over 4}\hskip1mm e^{-|t|/\tau}\cos\omega_0 t \hskip5mm \Rightarrow 
\label{a42}
\end{equation} 
$$
\Rightarrow {\langle g(0)g(t)\rangle\over g_0^2}=e^{-|t|/\tau}\cos\omega_0 t
$$
 We introduce phenomenologically the transverse relaxation time $\tau$. 
 For noise power spectrum we get
\begin{equation}
{\cal N}(\nu)={u_{10}^2\tau\over N_b}\hskip1mm \bigg [
{1\over 1+(\omega_0+\nu)^2\tau^2} + {1\over 1+(\omega_0-\nu)^2\tau^2}
\bigg ]
\label{a43}
\end{equation}      
The root-mean-square of polarimetric noise is defined by following relationship 
 \begin{equation}
 \langle \delta u^2\rangle={1\over 2\pi}\int  {\cal N}(\nu)d\nu=  {u_{10}^2\over  N_b}
\label{a44}
\end{equation} 
 In the similar way one can calculate the power spectrum of polarimetric noise Eqs. (\ref{a19},\ref{a29},\ref{a31})  in presence of AB. Using the relationship Eq. (\ref{a37}) for correlation function and Eqs. (\ref{a19},\ref{a29},\ref{a31}), one can obtain:  

\begin{equation}
             \langle u(0)u(t)\rangle=
             u_{10}^2{e^{-|t|/\tau}\cos\omega_0t \over N_b}\bigg [
             1+2\sqrt{W_t\over W}\exp\bigg [-{k^2\Theta^2\over 4Q^2}\bigg ]+ {W_t\over W}{1\over 2}\bigg (
                          1+\exp \bigg [-{k^2\Theta^2\over Q^2}\bigg ]\bigg )
             \bigg ]
               \label{a45}
             \end{equation}
     
 If $2\pi/k=1\mu$m, $\Theta\sim 0.1$ and $1/Q\sim 30\mu$m, the exponential factors can be omitted and  simplified expressions for  correlation function and noise power spectrum take the form
 
 \begin{equation}
              \langle u(0)u(t)\rangle= 
              {u_{10}^2 \over N_b}\bigg [
              1+ {W_t\over 2W}\bigg ]e^{-|t|/\tau}\cos\omega_0t
                \label{a46}
              \end{equation}

\begin{equation}
{\cal N}(\nu)={u_{10}^2\tau\over N_b}\hskip1mm \bigg [
              1+ {W_t\over 2W}\bigg ] \bigg [
{1\over 1+(\omega_0+\nu)^2\tau^2} + {1\over 1+(\omega_0-\nu)^2\tau^2}
\bigg ]
\label{a47}
\end{equation} 
It is seen from Eq. (\ref{a47}) that if $W_t\sim W$,  switching of  additional beam leads to 50$\%$ increasing of the noise power  and seems can  be    easily observed.
Note once again that we suppose complete overlapping of the probe and AB which may be not the case. For this reason the contribution of AB to noise power spectrum in real experiment may be less than predicted by Eqs. (\ref{a46}, \ref{a47}).

\section*{Conclusion}

We develop a single scattering theory for the purposes of optical spin noise spectroscopy. For introduced model of polarimetric detector we calculate its output signal produced by Gaussian beam probing the sample with specified gyrotropy and show that this signal is proportional to the sample's gyrotropy averaged over irradiated volume.  We calculate the signal produced by additional probe beam tilted with  respect to the main one and show that this signal is proportional to the Fourier transformation of gyrotropy calculated for the differential wave vector of  the main and additional beam. The noise signals within the model of independent paramagnetic centers are also calculated.

\newpage


\end{document}